\documentclass[prb,aps, superscriptaddress,twocolumn]{revtex4-2}
\usepackage{amsmath,amssymb}
\usepackage{graphicx}
\usepackage{wasysym}
\usepackage{amsfonts}
\usepackage{bm}
\usepackage{enumerate}
\usepackage{color}
\usepackage[resetlabels]{multibib}
\usepackage{epstopdf}
\usepackage{latexsym}
\usepackage[breaklinks,colorlinks = true,linkcolor = red,urlcolor=cyan,citecolor=red]{hyperref}
\usepackage[caption=false,singlelinecheck=false]{subfig}
\usepackage[normalem]{ulem}
\usepackage{chemformula}

\usepackage{times}
\newcommand{\bea}{\begin{eqnarray}}
\newcommand{\eea}{\end{eqnarray}}
\newcommand{\be}{\begin{eqnarray}}
\newcommand{\ee}{\end{eqnarray}}
\newcommand{\bw}{\begin{widetext}}
\newcommand{\ew}{\end{widetext}}

\newcommand{\tcb}[1]{\textcolor{blue}{#1}}

\begin{document}
\title{Reallocation of Nonlocal Entanglement in Incommensurate Cold Atom Arrays}
\author{Jemin Park}
\email{physjmp@kaist.ac.kr}
\affiliation{Department of Physics, Korea Advanced Institute of Science and Technology, Daejeon, 34141, Korea}
\author{Junmo Jeon}
\email{junmo1996@kaist.ac.kr}
\affiliation{Department of Physics, Korea Advanced Institute of Science and Technology, Daejeon, 34141, Korea}
\author{SungBin Lee}
\email{sungbin@kaist.ac.kr}
\affiliation{Department of Physics, Korea Advanced Institute of Science and Technology, Daejeon, 34141, Korea}
\date{\today}
\begin{abstract}
Cold atom arrays in optical lattices offer a highly tunable platform for exploring complex quantum phenomena that are difficult to realize in conventional materials. Here, we investigate the emergence of controllable long-range quantum correlations in a simulated twisted bilayer structure with fermionic cold atoms. By exploiting the incommensurate nature of the twisted bilayer, we observe a significant enhancement of long-range susceptibility, suggesting the formation of stable entangled states between spatially distant localized spins. We further show that the tunability of the interlayer coupling in terms of driving fields enables us to manipulate these entangled states without deformation of lattice structure and extra doping. Our findings provide a pathway for overcoming challenges in establishing strong correlations across distant sites, highlighting the potential of optical lattices as a versatile platform for advanced quantum technologies.
\end{abstract}


\maketitle

\textit{\tcb{Introduction---}}Experimental advancements in synthetic quantum systems, such as cold atom systems and photonic crystals, have significantly expanded the scope of the exploration of quantum phenomena that were previously inaccessible in traditional condensed matter systems\cite{jaksch2005cold, takamoto2005optical, morsch2006dynamics, esslinger2010fermi, eckardt2017colloquium, gross2017quantum, lopez2003materials, joannopoulos1997photonic}. These synthetic platforms offer exceptional experimental control, enabling precise manipulation of parameters such as geometry, interaction strength, and dynamics across extended parameter regimes. Cold atom systems, in particular, provide an excellent platform for studying many-body physics, with previous studies focusing on phenomena such as Bose-Einstein condensates, superfluid-Mott insulator transition, non equilibrium quantum dynamics and superconductivity\cite{morsch2006dynamics, bakr2010probing, greiner2002quantum, pichler2010nonequilibrium, denschlag2002bose, chin2006evidence, le2009superconductivity}. The ability to control quantum correlations in these systems has brought cold atom platforms to the forefront of emerging quantum technologies, including quantum computing, simulation, and sensing\cite{deutsch2000quantum, brennen1999quantum, gross2017quantum, ladd2010quantum, simon2011quantum, struck2011quantum, degen2017quantum, ranjit2016zeptonewton, sorrentino2009quantum}.

One of the key challenges to realize quantum technologies in conventional systems is the ability to establish and control quantum correlations over long distances between atoms. In general crystalline structures, however,  interactions are primarily local, and most well-known long-range interactions are uniformly decaying with spatial distance. Thus, the influence of interactions diminishes significantly as the separation between target atoms increases. Consequently, correlation strength typically exhibits rapid falloff, often following an exponential or power law decay with distance. However, unlike natural materials, optical lattices offer the advantage of easily tunable lattice structures, making them an ideal platform for exploring cases where the traditional decay of long-range correlations does not apply. Particularly, there has been recently growing interest in expanding optical lattice platforms beyond simple crystalline structures to more complex ones, such as quasicrystals and twisted multilayer systems\cite{PhysRevLett.122.110404, gonzalez2019cold, meng2023atomic,yu2024observing, sanchez2005bose}. These structures offer the potential for novel behaviors, as the wavefunctions in such lattices do not follow the simple Bloch functions of conventional crystals\cite{anderson1958absence, PhysRevB.35.1020}. This opens the door to fundamentally different physical phenomena, particularly those mediated by magnetic interactions of cold atoms, which remain largely unexplored yet could have profound implications for the development of quantum technologies\cite{jeon2025hidden}.

Another major challenge in experimentally implementing and controlling advanced technologies lies in the difficulty of tuning interaction strengths and properties in natural materials. In most conventional systems, interactions are primarily determined by the electronic structure, often dictated by the Fermi surface. As a result, modifying long-range interactions and correlations within a given material typically requires altering the Fermi level through doping or lattice deformation. However, in metallic systems, even doping provides limited control over the Fermi level. In contrast, optical lattice platforms allow precise control over cold atom hopping energies at the microscopic level.\cite{ibanez2013tight, gonzalez2019cold,jaksch1998cold} This enables direct manipulation of the energy spectrum and wavefunctions while preserving the lattice structure, making optical lattices a highly tunable and experimentally accessible platform for quantum technologies.

\begin{figure}[h]
\hfill
\subfloat[]{\includegraphics[width=0.30\textwidth]{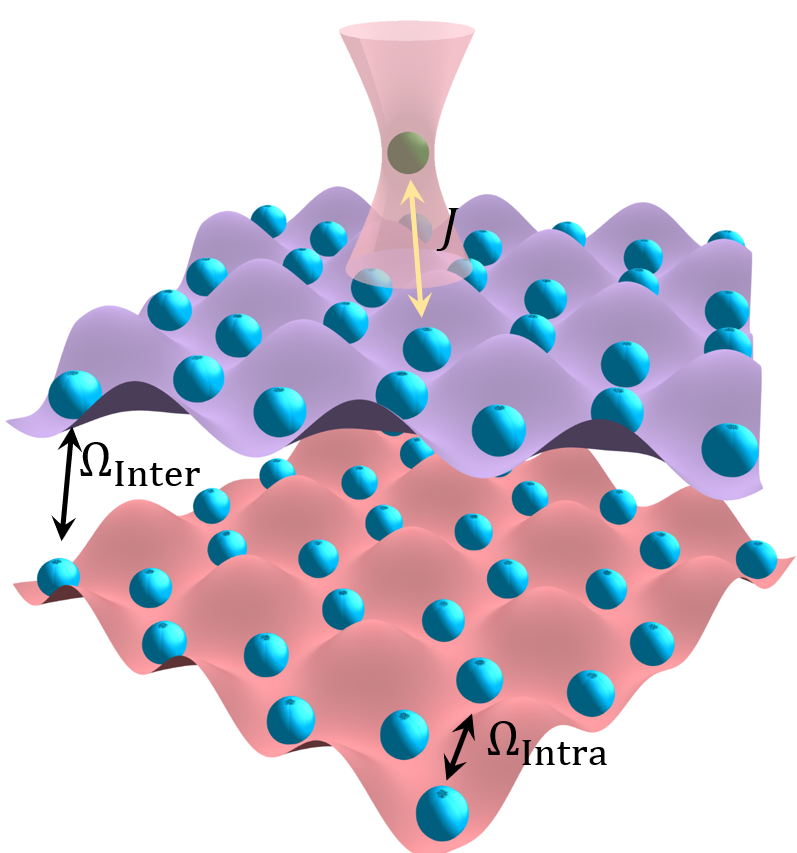}}
\hfill
\subfloat[]{\includegraphics[width=0.18\textwidth]{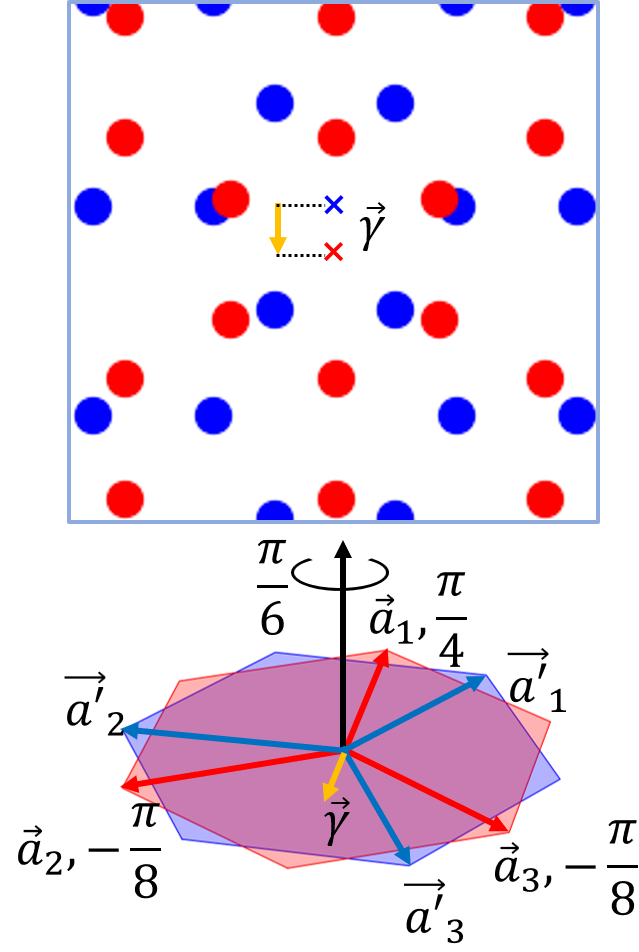}}
\caption{(a) Schematic illustration of the cold atom arrays in a bilayer optical lattice system, where each layer is formed by a honeycomb lattice generated by the interplay between three laser fields. The potential landscapes of these layers are state-dependent, and the interlayer coupling strength, $\Omega_{\text{inter}}$ can be tuned via an external driving field. Local magnetic impurities can be introduced by positioning magnetic atoms using optical tweezers. In this work, we consider Kondo interaction between a cold atom and a magnetic impurity with coupling strength $J$. (b) Structure of a $30^\circ$-twisted shifted bilayer honeycomb lattice. The red and blue dots represent the atomic positions of each layer, while the `x' marks the rotation center of each lattice. $\vec{a}_i$ and $\vec{a'_i}$ denote the laser wavevectors, which are related by a $30^\circ$ rotation. The red layer is shifted by a quarter of $\vec{a}_1$ with proper set of phase shifts in laser fields. See the main text for details.
}
{\label{fig:lattice}}
\end{figure}

In this Letter, we show that the long-range entanglement is not only stable but controllable in the incommensurately twisted bilayer optical lattice. By exemplifying 30$^\circ$ twisted bilayer honeycomb optical lattice with some translational shift (Refer Fig.\ref{fig:lattice}), we unveil that the incommensurability between two honeycomb optical lattices gives rise to the anomalous enhancement of long-range correlations between arbitrarily distant localized spins. Furthermore, we figure out that the characteristics of the long-range correlations can be experimentally changeable in terms of tunable interlayer couplings. Interestingly, we demonstrate that by tuning the interlayer coupling strength, it is possible to reallocate long-range entanglement from one pair of atoms to another without any additional doping or deformation of the lattice structure. This provides the 
non-destructive way to manipulate long-range entangled states, which are stable under unavoidable thermal fluctuation. Such findings can significantly contribute to advancing cutting-edge quantum communication technologies and exploring exotic many-body physics. 

\textit{\tcb{Twisted bilayer optical lattice---}} 
The bilayer optical lattice structure could be constructed with two key ingredients. One is the polarization depending atomic states with three different wavevectors of the lasers forming independent interference pattern. The other one is the external driving field which alters the atomic state. The lasers generate polarization-dependent potentials which depend on the atomic states, such as angular momentum. As a result, each atomic state is confined within the corresponding laser potential. Moreover, one can invert one atomic state to another in terms of the external driving field. This realizes the effective interlayer hopping between two optical lattices. \cite{gonzalez2019cold,meng2023atomic}

Let us consider a twisted and shifted bilayer honeycomb optical lattices as a concerete example of incommensurate system. We use two sets of lasers, each generating an independent honeycomb optical lattice. Each honeycomb layer consists of lasers with momenta in three different directions, say $\vec{a}_i$ and $\vec{a}_i'$, respectively. By rotating the momenta of one set of lasers with respect to the other set of lasers, we can simulate 30$^\circ$ twisted bilayer honeycomb system. In addition, we adopt phase modulations of the lasers to induce a shift of one-quarter of the primitive lattice vector to one layer.

Fig.\ref{fig:lattice} illustrate the schematic figure of shifted 30$^\circ$ twisted bilayer honeycomb optical lattice. Here, the red and blue colors represent each layer. By introducing the phase shifts of $\pi/4$, $-\pi/8$, and $-\pi/8$ to the lasers forming red layer, the red layer is shifted by $\vec{\gamma}=-\frac{1}{4}\vec{a}_1$. As one layer is shifted by a quarter of the primitive lattice vector, the symmetry of the resulting structure is reduced to the mirror symmetry along the direction of the shift. In these lattices, cold atoms move between different sites, with interlayer and intralayer hopping magnitudes denoted as $\Omega_\text{Inter}$ and $\Omega_\text{Intra}$, respectively. We also introduce magnetic impurities to realize Kondo coupling $J$, positioning them precisely using optical tweezers.
The interaction between the cold atoms and magnetic impurities will be discussed in the next section.

Now, we simulate fermionic particles in the shifted and twisted bilayer optical lattice. Let us consider the pseudo spin-1/2 Alkaline-Earth cold atoms whose dynamics is governed by the Hamiltonian given by, \cite{gonzalez2019cold},
\begin{equation}\label{eq:Hamiltonian}
H =  \sum_{i,j, i \neq j, \sigma}J_{ij}{c_{i,\sigma}^\dag}{c_{j,\sigma}},
\end{equation}
where $c_{i,\sigma}^\dag$ and $c_{i,\sigma}$ are the creation and annihilation operators spin $\sigma$ at the $i$ site.
The hopping integral between site $i$ and site $j$ with displacement $\vec{R}_{ij}$ can be represented as,
\begin{align}\label{eq:transfer_integral}
J_{ij} = \Omega_{\text{intra/inter}} e^{-|\vec{R}_{ij}|^2/4L^2_0},
\end{align}
where $ L_0 $ is the size of the ground-state wavefunction, determined by the mass of the cold atom and the trapping frequency. $ \Omega_{\text{intra}} $ and $ \Omega_{\text{inter}} $ represent the intra- and inter-layer hopping strengths, respectively, with $ \Omega_{\text{inter}} $ being independently tunable from $\Omega_{\text{intra}}$ by external driving field.
These tunabilities enable the modulation of the hopping integral beyond real materials. Here, we define the strength of incommensurability of the system as  $\alpha=\vert\Omega_{\text{inter}}/\Omega_{\text{intra}}\vert$.

We now focus on the strong interlayer coupling regime, as when $\alpha\ll 1$, the system reverts to two independent honeycomb lattices. Let us consider $\alpha\gtrsim 5$, where $\Omega_{\text{intra}}$ is on the order of $1\mu K$ scale. We set $L_0 = 0.1565a$, where $a$ represents the lattice constant of the honeycomb lattice. Note that $a$ is determined by the wavelength of the laser used, and it allows for the tuning of $\Omega_{\text{intra}}$. We assume $a$ is on the order of $0.1\sim 1 \mu m$. Then, the nearest neighbor hopping is on the order of $10\sim 100 nK$ scale.\cite{lewenstein2012ultracold} 

\begin{figure}[h]
\includegraphics[width=0.50\textwidth]{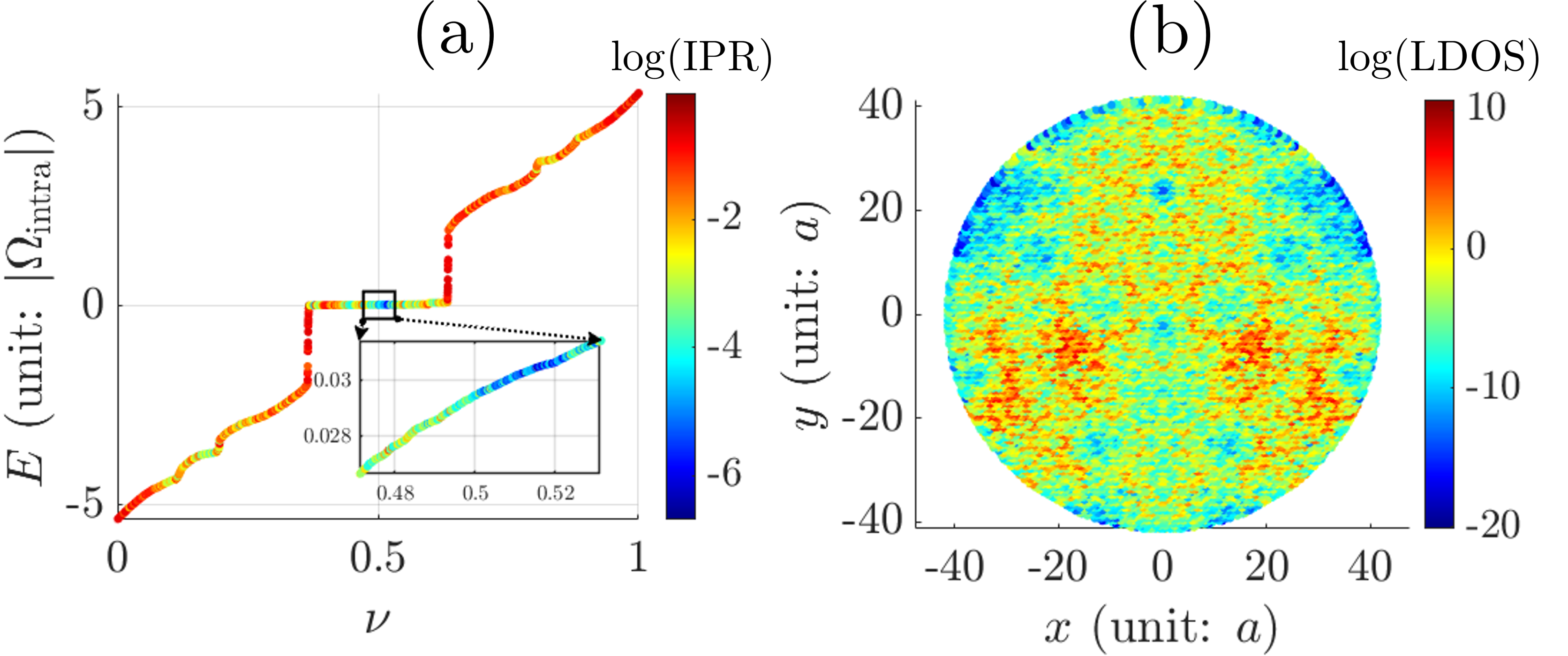}
\caption{(a) Energy spectrum of strong interlayer coupling regime. Here, $\Omega_{\text{intra}}=-2.7$ and $\Omega_{\text{inter}}=14.4$. The colors represent the IPR of the eigenstates in log-scale. The inset highlights around the half-filling regime, where $\vert E\vert\ll\vert\Omega_{\text{intra}}\vert$. (b) Local density of states at the filling fraction, $\nu=0.5049$ in log-scale that tends to concentrate over long distances between sites with similar surrounding pattern. The number of sites is $24932$.}
{\label{fig:wave function}}
\end{figure}

When the interlayer coupling is much stronger than the intralayer coupling, the energy spectrum can be categorized into two regimes: (1) negligible intralayer coupling and (2) significant intralayer coupling effects. Specifically, the intralayer coupling effect is negligible in eigenstates whose energy $E$ such that $\vert E \vert \gg \vert \Omega_{\text{intra}} \vert$. In this regime, the wavefunctions are insensitive to the lattice geometry over large distances and are primarily localized to the local geometry. While, for eigenstates with energies near or smaller than the intralayer coupling strength, a more pronounced interplay between intra- and interlayer dynamics would emerge. In detail, the eigenstates in this regime would be strongly concentrated at some widely separated positions, sharing a common surrounding pattern (see Fig.\ref{fig:wave function} (b)). Note that by fixing the filling fraction, $\nu=N_a/2N$ of pseudo-spin-1/2 cold atoms, we can focus on a specific energy range to explore physical properties. Here, $N_a$ and $N$ denote the numbers of cold atoms and optical lattice sites, respectively.

Fig.\ref{fig:wave function} shows the energy spectrum and local density of state (LDOS) for strong interlayer coupling regime. Fig.\ref{fig:wave function} (a) presents the energy spectrum, highlighting the inverse participation ratio (IPR) of each eigenstate, defined as $\sum_i\vert \psi_n(i)\vert^4$, where $n$ denotes the energy index and $\psi_n(i)$ represents the wavefunction of the $n$th eigenstate at site $i$. In the regime of $\vert E \vert \ll \vert \Omega_{\text{intra}} \vert$, the values of IPR are remarkably small due to the interplay between intralayer and interlayer hoppings (see the inset of Fig.\ref{fig:wave function} (a)). In contrast, almost all states become localized in the regime where $\vert E\vert\gg \vert\Omega_{\text{intra}}\vert$. Fig.\ref{fig:wave function} (b) illustrates the LDOS, defined as $\mathrm{LDOS}(i,E)=-\frac{1}{\pi}\mathrm{Im}G_{ii}^r(E)$ near the half-filling regime ($\nu=0.5049$). Here, $G_{ii}^r(E)$ is retarded Green's function of the Hamiltonian in Eq.\eqref{eq:Hamiltonian} at site $i$. Notably, the LDOS in Fig.\ref{fig:wave function} (b) is concentrated at mirror-symmetric positions with a shared local pattern. This suggests that the indirect interactions mediated by these states are enhanced between such widely separated locations, as discussed below.

\textit{\tcb{Anomalously enhanced long-range susceptibility---}} Now, we explore long-range interactions focusing on the strong interlayer coupling regime. We investigate the response to a stimulus at a given site by examining the static susceptibility, which captures the influence of perturbations over long distances. For concrete argument, we consider long-range correlations between localized pseudo-spin half atoms, which are placed at certain sites by using optical tweezers. The localized magnetic atoms, placed via optical tweezers, interact with the pseudo-spin moments of the cold atoms at their respective positions through the following local Kondo coupling: $H_\mathrm{local} = \frac{J}{2} \sum_{i \in \mathcal{D}} \vec{S}_i \cdot c_{i}^\dag \vec{\sigma} c_i$, where $J$ is the local coupling strength, $\mathcal{D}$ is the set of positions of the magnetic impurities, $\vec{\sigma}$ represents the Pauli matrices, and $\vec{S}_i$ is the magnetic moment of the impurity at site $i$. By integrating out the spin degrees of freedom of the cold atoms, the effective interaction between the magnetic impurities can be written as: $\mathcal{H} = J^2\sum_{i\neq j}\chi_{ij}\vec{S}_i \cdot \vec{S}_j$, where $\chi_{ij}$ is the static susceptibility.

The static susceptibility between sites $i$ and $j$, $\chi_{ij}$ mediated by itinerant particles is given by $\chi_{ij}=-\frac{1}{2\pi} \int_{-\infty}^{E_F} \text{Im}[G_{ij}^r(E)G_{ji}^r(E)]dE$, where $E_F$ is the Fermi energy. 
In energy basis, $\chi_{ij}$ is written as,
\begin{multline}\label{eq:susceptibility_E}
\chi_{ij}=\frac{1}{4}\sum_{m,n,{E_m}\neq{E_n}}\text{Re}[\psi_m(i)\psi_m(j)^*\psi_n(j)\psi_n(i)^*]\\
\times \frac{n_F(E_n-E_F)-n_F(E_m-E_F)}{E_n-E_m}.
\end{multline}
Here, $m$ and $n$ are energy indices. $\psi_m(i)$ is the wavefunction of the eigenstate whose energy $E_m$ at site $i$. $n_F(E)=(1+\exp{(E/{k_BT})})^{-1}$ is the Fermi-Dirac distribution, where $T$ is temperature. The wavefunctions near the Fermi energy mainly contribute to the susceptibility at low temperature. To capture the interplay between intra- and interlayer couplings, we set the filling fraction, $\nu$ to $0.5049$, ensuring that the Fermi level lies within the energy window $\vert E \vert \lessapprox \vert \Omega_{\text{intra}} \vert$.

Remind that the static susceptibility strength uniformly decays with distance in either exponential or power-law form in conventional periodic crystals, where the eigenstates are either exponentially localized or uniformly extended. Thus, in conventional crystalline systems, interactions over long distances are negligible\cite{PhysRev.96.99, RevModPhys.34.681}. However, this is no longer the case in our system which possesses the exotic quantum states shown in Fig.\ref{fig:wave function}.

\begin{figure}[h]
\includegraphics[width=0.45\textwidth]{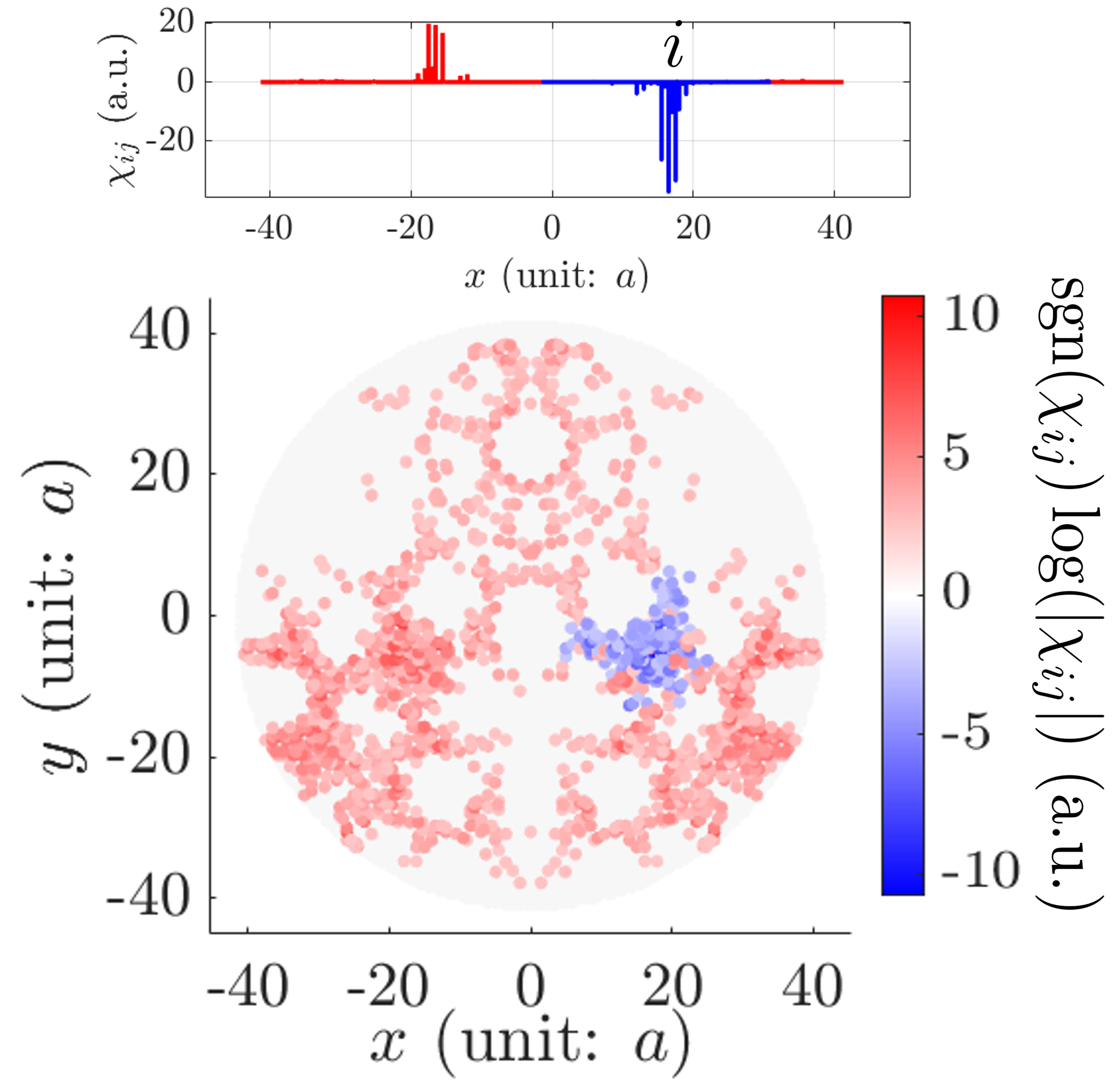}
\caption {Anomalously enhanced long-range static susceptibility, $\chi_{ij}$, with a fixed site $i$. Red (blue) color represents the positive (negative) susceptibility, respectively. The susceptibility is anomalously enhanced even for long distance, $\vert\vec{R}_{ij}\vert\gtrsim 40a$. Here, the susceptibility is expressed in arbitrary units (a.u.), and all values with $\vert \chi_{ij}\vert<1$ are omitted. The fixed site $i$ is depicted in the top panel. We set ${k_B}T = 10^{-4}$, $\Omega_{\text{intra}}=-2.7$, $\Omega_{\text{inter}}=14.4$ and $\nu=0.5049$. The number of sites is 24932. 
}
{\label{fig:RKKY_all}}
\end{figure}

Fig.\ref{fig:RKKY_all} shows the anomalous increase in long-range susceptibility for a given $i$ site. The blue (red) color corresponds to positive (negative) values, respectively. In particular, $\chi_{ij} > 0$ across the mirror plane. The top panel of Fig.\ref{fig:RKKY_all} emphasizes the anomalously enhanced susceptibility at long distances. The susceptibility peaks appear at the mirror-symmetric points and closely correspond to the locations of the predominant wavefunctions (compare Figs.\ref{fig:wave function} (b) and \ref{fig:RKKY_all}). Thus, $\chi_{ij}$ is selectively enhanced in specific location pairs determined by states near the Fermi level, rather than appearing randomly.

\begin{figure}[h]
\includegraphics[width=0.4\textwidth]{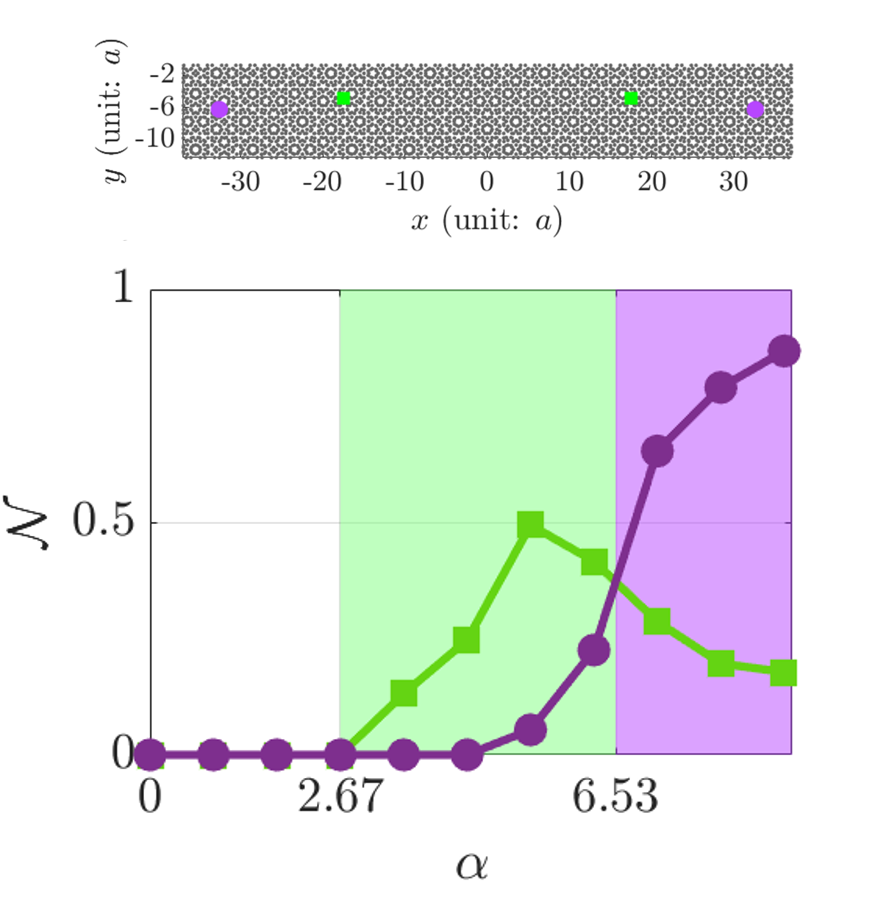}
\caption {Stable entanglement reallocation between two distinct spin-1/2 pairs as a function of interlayer coupling strength. The top panel is shown for emphasizing two distinct pairs of localized magnetic impurities that are marked by green squares and violet circles, respectively. Each pair consists of mirror symmetric positions. The green square and violet circle curves in the bottom panel show the logarithmic negativity for each colored pair of impurities as the function of strength of incommensurability, $\alpha$. Here, $\Omega_{\text{intra}}=-2.7$ and $\Omega_{\text{inter}}\ge 0$. We set $k_B T=10^{-4}$, $\nu=0.5049$, $N=24932$ and $\vert J\vert=2\times10^{-2.5}$. The entanglement of these two pairs of impurities exhibits three different features depending on $\alpha$: ($0\le\alpha<2.67$) No entanglement is observed at finite temperature condition. ($2.67\le\alpha<6.53$) The violet pair admits only negligible entanglement, but green pair shows large entanglement. ($\alpha\ge 6.53$) Notably, the violet pair is rapidly entangled, while the entanglement of green pair diminishes as the interlayer coupling increases.
}
{\label{fig:log_negativity}}
\end{figure}

\textit{\tcb{Tunable long-range entanglement---}} In many condensed matter experimental settings, reliably establishing and controlling long-range quantum entanglement is an important but challenging task. First, since the interactions in real materials typically diminish over distances beyond the mesoscopic scale, requiring either numerous intermediary spins or very low temperature conditions to form such long-range entanglement. Second, despite indirect long-range interactions and the resulting quantum correlations can be changed by temperature and filling fraction, the range of control over these effects is limited, especially, in metallic system. Moreover, the changes in external parameters like temperature affect the system, making it difficult to achieve precise control between qubits. Hence, we have a significant limitation for fine-tuning stable entanglement in conventional systems.

We emphasize that the selectively enhanced long-range interactions in the incommenusrate optical lattice enable us to stably control the long-range entanglement in terms of the driving fields. Note that tuning driving fields only changes the interlayer coupling without deforming lattice structure or altering filling fraction. Nevertheless, tuning driving fields directly changes the microscopic Hamiltonian, Eq.\eqref{eq:Hamiltonian}, and the characteristics of wavefunctions near the Fermi energy for fixed filling fraction. This greatly expands experimental applications, enabling the exploration of strongly correlated phases and quantum resources such as entanglement and coherence between distant pairs.

To clarify the stable reallocation of entanglement over long distances, we investigate the entanglement between two distinct widely separated pairs of magnetic impurities at the finite temperature. Specifically, we place impurity spins at four sites, which are divided into two pairs: one pair represented by green squares and the other by violet circles in the top panel of Fig.\ref{fig:log_negativity}. We denote each pair by subscripts $\text{`g'}$ and $\text{`v'}$, respectively. We then examine the entanglement between each pair of sites as a function of the interlayer coupling strength. Note that under $T>0$, the quantum state is described by a mixed state, $\rho=\frac{e^{-\mathcal{H}/k_BT}}{\text{Tr}(e^{-\mathcal{H}/k_BT})}$. To quantify the entanglement, we adopt negativity for each reduced density matrices, $\rho_{\text{g(v)}}=\text{Tr}_\text{v(g)}\rho$, given by $\mathcal{N}_{\text{g(v)}}=\log_2 \lVert \rho_{\text{g(v)}}^{\Gamma_A} \rVert$\cite{peres1996separability}. Here, $\Gamma_A$ stands for partial transpose with respect to subsystem $A$ and $\lVert \rho \rVert = \text{Tr}\sqrt{\rho^{\dag} \rho}$. Since the negativity is zero for separable states, a nonzero value of logarithmic negativity serves as a sufficient condition for entanglement.

To show the reallocation of entanglement in terms of $\alpha$, we consider the case where the interaction between the green and violet sites is negligible compared to the interactions within each pair. This assumption can be realized through precise impurity placement using optical tweezers, as indicated by the highly selective susceptibility distribution shown in Fig.\ref{fig:RKKY_all}. However, we emphasize that our results can be extended to situations where a significant coupling exists between the green and violet sites. In the case where the interaction between the two pairs is negligible, the Hamiltonian is decoupled, and each pair of green and violet sites has entanglement depending on the long-range interaction within pair. Specifically, when the long-range interaction between two sites is $J(\alpha)>0$ at low temperature, their entanglement negativity is given by, 
\begin{align}
    \label{neganega}
    &\mathcal{N}(\alpha)=\log_2\left( \frac{2}{3 e^{-\beta J(\alpha)}+1}\right),
\end{align}
that is a monotonic function of $J(\alpha)$. Thus, non-monotonic behavior of $J$ as a function of $\alpha$ can lead to the reallocation of entanglement as we will show.

Given filling fraction and finite temperature, Fig.\ref{fig:log_negativity} shows the logarithmic negativity for each pair as the functions of strength of incommensurability, $\alpha$. Interestingly, as interlayer coupling strength increases, the negativity shows three different characteristics. First, for small values of $\alpha$ (see the white region in Fig.\ref{fig:log_negativity}), the system is approximated as two independent honeycomb lattices. Thus, both pairs are decoupled under the finite temperature condition. Next, as the interlayer coupling increases (see the green region in Fig.\ref{fig:log_negativity}), the green pair begins to become entangled, which reveals the thermally stable entanglement originated from the strong interlayer coupling. While, the entanglement of violet pair is still low. This indicates the selective formation of entanglement. Lastly, as the interlayer coupling becomes significantly stronger (see the violet region in Fig.\ref{fig:log_negativity}), the entanglement of the violet pair increases rapidly, while that of the green pair decreases and eventually falls below that of the violet pair.
This phenomenon arises because the
concentration of probability density near the Fermi energy moves from green sites to violet ones as $\alpha$ increases.
This demonstrates the reallocation of long-range entanglement between atom pairs by tuning the interlayer coupling strength through the driving field intensity.

\tcb{\textit{Conclusion---}} We have investigated that incommensurately twisted bilayer optical lattices provide a robust platform for realizing and manipulating long-range quantum entanglement. The interplay between strong interlayer coupling and incommensurability leads to anomalous enhancements of long-range susceptibility, enabling stable entanglement reallocation between distant atomic pairs. Our findings show that long-range entanglement can be selectively tuned and redistributed via interlayer coupling strength. This tuning is experimentally achievable through driving fields. Notably, it requires no structural deformations or additional dopings. These features make the system a highly controllable platform for quantum technologies.

Our analysis is broadly applicable to incommensurate optical lattices, where spatially inhomogeneous wave functions arise. In these systems, sites separated by long distances can share similar surrounding patterns, leading to a highly concentrated local density of states between them. As a result, strong entanglement emerges between magnetic impurities at distant sites and by tuning the hopping parameters, entanglement allocation can be controlled, similar the scenario presented here.
Our study opens the possibility of engineering long-range entangled states in cold atom systems, with applications in quantum communication and many-body physics. By utilizing incommensurate optical lattices, it highlights a new path for exploring unconventional correlations and novel quantum phases.


\begin{acknowledgments}
\noindent	
{\em Acknowledgments.---}
We would like to thank Kiryang Kwon for his discussions and insightful comments on the cold atom system. This research was supported by National Research Foudation Grant (2021R1A2C109306013).

\end{acknowledgments}

\appendix

\bibliography{biblio}


\end{document}